\documentclass{elsart} %\documentclass{my_text} %\documentclass{article}
\usepackage{epsfig}
\usepackage{amssymb}  %\usepackage{amsmath, amssymb}
\usepackage{axodraw}

\def\xboxt#1{ \begin{tabular}{|l|} \hline #1 \\ \hline \end{tabular} }

\begin{document}

\begin{frontmatter} % Title, authors and addresses
%\begin{topmatter}

% and the from \ead[url] for the home page:
%\title{ALHEP manual \thanksref{thanks1}}
%\title{ALHEP manual}
\title{ALHEP symbolic algebra program for high-energy physics}


\author{V. Makarenko \thanksref{INTAS}}

\ead{makarenko@hep.by}
\thanks[INTAS] { Supported by INTAS YS Grant no. 05-112-5429 }
%\ead[url]{http://cern.ch/~makarenko/alhep, http://www.hep.by/alhep}
%\email{makarenko@hep.by}

\address{NC PHEP BSU, 153 Bogdanovicha str., 220040 Minsk, Belarus}
%\institution{NC PHEP BSU}
%\address{153 Bogdanovicha str., 220040 Minsk, Belarus}

\begin{abstract}
ALHEP is the symbolic algebra program for high-energy physics.
It deals with amplitudes calculation, matrix element squaring,
Wick theorem, dimensional regularization, tensor reduction of loop integrals
and simplification of final expressions.
The program output includes:
Fortran code for differential cross section,
Mathematica files to view results and intermediate steps
and TeX source for Feynman diagrams.
The PYTHIA interface is available.

The project website {
\verb|http://www.hep.by/alhep|
%\verb|http://cern.ch/~makarenko/alhep|
}
contains up-to-date executables, manual and script examples.
\end{abstract}

%\begin{keyword}

%\PACS 
%\end{keyword}

\end{frontmatter}
%\begin{topmatter}

%\runauthor{V. Makarenko}
%\runtitle{ALHEP manual (draft)}

\section{Introduction}
\label{sec_INTRO}
The analytical calculations in high-energy physics
are mostly impossible without a powerful computing tool.
The big variety of packages is commonly used
\cite{CompHEP,SANC,GRACE,MadGraph,OMega,FormCalc,FeynCalc,Amegic,AlpGen,HELAC,xloops,aiTALC,MINCER,DIANA}.
Some are general-purpose symbolic algebra programs
with specific HEP-related plug-ins (REDUCE \cite{REDUCE}, Mathematica \cite{Mathematica}),
some are designed especially for particle physics
(CompHEP \cite{CompHEP}, SANC \cite{SANC}, GRACE \cite{GRACE} etc.)
and some are created for specific interaction class or specific task.
Many of them uses external symbolic algebra core (Form \cite{FORM}, MathLink \cite{Mathematica}).
They can deal with matrix elements squaring (FeynCalc \cite{FeynCalc})
or calculate helicity amplitudes directly (MadGraph \cite{MadGraph}, CompHEP \cite{CompHEP}, O'Mega \cite{OMega}).
Some packages provide numerical calculations,
some require external Monte-Carlo generator to be linked.
Some programs contain also one-loop calculation routines
(FormCalc \cite{FormCalc}, GRACE \cite{GRACE}, SANC \cite{SANC}).
Nevertheless, there is no uniform program that meets all the user requirements.

Every calculation requires the program-independent check.
The optimal tactics is the simultaneous usage of two (or more)
different symbolic algebra packages.

ALHEP is a symbolic algebra program for performing the way
from Standard Model Lagrangian
to amplitude or squared matrix element for the specified scattering process.
It can also be useful for loop diagrams analysis.
The basic features are:
\vspace{-5pt}
\begin{itemize}
\item Diagrams generation using Wick theorem and SM Lagrangian.
\item Amplitude calculation or matrix element squaring.
\item Bondarev functions method for traces calculation.
\item Tensor reduction of loop integrals.
\item Dimensional regularization scheme.
\item Generation of \verb|Fortran| procedures for numerical analysis
(PYTHIA \cite{PYTHIA} and LoopTools \cite{LoopTools} interfaces are implemented).
\end{itemize}

The current ALHEP version
%is declared as draft and 
have several implementation restrictions,
that will be lifted in future.
The following features are in progress of implementation:
\vspace{-5pt}
\begin{itemize}
\item Bremsstrahlung part of radiative correction (the integration over real photon phase space).
\item Complete one-loop renormalization scheme including renormalization constants derivation.
\item Arbitrary Lagrangian assignment.
\end{itemize}%\end{enumerate}
After these methods implementation the complete one-loop analysis will be available.
%At present ALHEP can be useful for amplitudes and box-like corrections computation.
Please refer to project website for program updates.

ALHEP website \verb|http://www.hep.by/alhep|
contains the up-to-date executables (for both Linux \& Win32 platforms),
manual and script examples.
The mirror at \verb|http://cern.ch/~makarenko/alhep| is also updated.

%This review starts form installation notes and script is organized as follows:
%...

%{\bf Disclaimer.}
%ALHEP is a young program and some
%of the declared features are implemented with restrictions or not tested well. There are still many work on its improvement

%\newpage
\section{ALHEP Review. Program Structure \label{sec_REVIEW}}
%{\bf Program Structure}

The ALHEP program internal structure can be outlined as follows:
\vspace{-5pt}
\begin{itemize}%\begin{enumerate}
\item The native symbolic algebra core.
\item Common algebra libraries:

Dirac matrices, tensor and spinor algebra, field operators \& particle wave functions zoo.

\item Specific HEP functions and libraries.

It include Feynman diagrams generation, trace calculations, helicity amplitudes method,
HEP-specific simplification procedures, tensor integrals reduction and others.
\item Interfaces to Mathematica, Fortran, TeX and internal I/O format.

Fortran code is used for further numerical analysis.
Mathematica code can be used for viewing any symbolic expression in program.
But no backward interface from Mathematica is currently implemented.
TeX output can be generated for Feynman diagrams view.
Internal I/O format is implemented for the most of symbolic expressions,
allowing save and restore calculation at intermediate steps.
\item Command script processor.

User interface is implemented in terms of command scripts.
The ALHEP script language have C-like syntax, variables, arithmetic operations, function calls.
All HEP-related tasks are implemented as build-in functions.
\end{itemize}%\end{enumerate

\subsection{Getting Started \label{sec_DEMO}}

To use ALHEP one should download the pre-compiled executable for appropriate platform (Linux and Win32 are available)
and write a control script to describe your task.
ALHEP program should be launched with the single argument: script file name to be invoked
The following steps are required to create a workspace:
\vspace{-5pt}
\begin{itemize}
\item Download ALHEP executables from project website: \verb|alhep.gz| (for Linux) or \verb|alhep.zip| (Win32).
Unpack executable, e.g.

\verb|gzip -d alhep.gz|

\item Download up-to date command list: \verb|ALHEPCommands.txt|.
The set of commands (or options) may be changed in future versions and this manual may be somewhat obsolete.
Please refer to \verb|ALHEPCommands.txt| file that always corresponds to the latest ALHEP version.
The last changes are outlined in \verb|RecentChanges.txt| file at website.

\item Create some working directory and compose command script file therein.
For example consider the $u \bar{u} \to W^+ W^- \gamma$ process with $\{-+\}$-helicities of initial quarks.
To calculate amplitude we create the following script file (call it \verb|"test.al"|):
\vspace{4pt}
\begin{verbatim}
SetKinematics(2, 3               // 2->3 process
  ,QUARK_U,"p\_1","e\_1"         // u
  ,-QUARK_U,"p\_2","e\_2"        // u-bar
  ,WBOZON,"f\_1","g\_1"          // W{+}
  ,PHOTON,"f\_0","g\_0"          // photon
  ,-WBOZON, "f\_2", "g\_2" );    // W{-}
SetDiagramPhysics(PHYS_SM_Q1GEN);// SM with 2 quarks only
SetMassCalcOrder(QUARK_U, 0);    // consider massless
SetMassCalcOrder(QUARK_D, 0);    // consider massless
diags = ComposeDiagrams(3);      // create diagrams, e^3
DrawDiagrams(diags, "res.tex",
             DD_SMALL|DD_SWAP_TALES, FILE_START);
SetFermionHelicity(1, -1);
SetFermionHelicity(2, 1);
SetParameter(PAR_TRACES_BONDEREV, 1);
ampl = CalcAmplitude(RetrieveME(diags));
ampl = KinArrange(ampl);         // arrange result
ampl = Minimize(ampl);           // minimize result
SaveNB("res.nb",ampl,"",FILE_START|FILE_CLOSE);
f = NewFortranFile("res.F", CODE_F77); 
CreateFortranProc(f, "UUWWA", ampl,
  CODE_IS_AMPLITUDE|CODE_COMPLEX16
  |CODE_CHECK_DENOMINATORS|CODE_PYTHIA_VECTORS);
\end{verbatim}
\vspace{4pt}
The \verb|SetKinematics()| function declares particles, momenta and polarization symbols.
The physics is declared with \verb|PHYS_SM_Q1GEN| option to restrict diagrams number.
The amplitude for all ($d,c,b$) internal quarks can be obtained from generated here
by quark mixing matrix replacement: $U^2_{ud} \to U^2_{ud}+U^2_{us}+U^2_{ub}$ (chiral limit).
Diagrams are created by \verb|ComposeDiagrams()| call.
\verb|CalcAmplitude()| function creates the symbolic value for process amplitude.
For detailed discussion of this example see sec. \ref{sec_EX1}.

\item Create simple batch file \verb|"run.me"| like:
\begin{verbatim}
~/alhep_bin_path/alhep test.al
\end{verbatim}
The ALHEP program creates some console output (current commands, scroll bars and some debugging data).
If it is not allowed one should redirect console output to file here.%in your batch script.
\end{itemize}

The batch execution with \verb|test.al| command file takes about 1 minute at 1.8GHz P4 processor.
The following files are created:
%\vspace{-5pt}
\begin{description}
\item[\tt res.nb:] Mathematica file containing symbolic expression of amplitude. Created by \verb|SaveNB()| function.
\item[\tt res.F:] F77 code for numerical analysis created by \verb|NewFortranFile()| and \verb|CreateFortranProc()| calls.
The library file \verb|alhep_lib.F| is required for code compilation and should be downloaded from project website.
See sec. \ref{sec_FORTRAN} for Fortran generation and compilation review.
\item[\tt res.tex:] TeX source for $19$ Feynman graphs generated. The AxoDraw \cite{AxoDraw} Latex package is used.
The \verb|res.tex| file should be included into your LaTeX document using \verb|\input res.tex| command.
The template document to include your diagrams can be found at program website.
See fig. \ref{fig_uuWWA} in sec. \ref{sec_EX1} for diagrams generated.
\item[\tt debug.nb:] Mathematica file with debugging information and some intermediate steps.
The amount of debugging information is declared in \verb|debug.ini| file in working directory.
See sec. \ref{sec_DEBUG} for further details.
\end{description}

See sec. \ref{sec_EXAMPLES} and project website for another examples.
It is convenient to use some example script as template and modify it for your purposes.
%The project examples are available at program website.
%One should also insert proper path to ALHEP executable in the example batch files.

\subsection{Calculation scheme \label{sec_SCHEME}}

The usual ALHEP script contains several steps:
% discussed here.
%Let's discuss the basic steps for the usual calculation using ALHEP.
%The following steps should be implemented in control script before program launching.

\begin{enumerate}
\item Initialization section

Declaration of process {\it kinematics}:
initial and final-state particles,
titles for particle momenta \& polarization vectors (\verb|SetKinematics()|).

{\it Physics model} definition.
The SM physics or part of SM Hamiltonian should be specified (\verb|SetDiagramPhysics()|).
The shorter Hamiltonian is selected, the faster is Wick theorem invocation.

{\it Polarization} declaration.
Every particle is considered as polarized with abstract polarization vector by default.
The specific helicity value can be set manually or particle can be marked as unpolarized.
The several ways of polarization involving are inplemented, see sec. \ref{sec_POL}
(\verb|SetPolarized()|, \verb|SetFermionHelicity()|, \verb|SetPhotonHelicity()|, ...).

Setting {\it mass-order} rules for specific particles (\verb|SetMassCalcOrder()|).
One can demand the massless calculation for light particle,
that greatly saves evaluation time.
One can also demand keeping particle mass with specific order $M^n$
and drop out the higher-order expressions like $M^{n+1}$.
It allows to consider the leading mass contribution without calculating precisely.

\item Diagrams generation

Feynman diagrams are generated for specific $e^n$ order using the {\it Wick theorem} algorithm (\verb|ComposeDiagrams()|)
%It may take much time for long Hamiltonian due to $n!$-complexity of the With theorem algorithm.
%There is also a {\it fast} diagram generation method, based on the common topology principles,
%but it may lead to erroneous results (is some cases) and may be used for debugging purposes only.
User may {\it draw diagrams} here (\verb|DrawDiagrams()|),
halt the program (\verb|Halt()|) and check out if diagrams are generated correctly.
%The diagrams are drawing in terms of AxoDraw \cite{AxoDraw} LaTeX package.

After the diagram set is generated one may {\it cut-off} not interesting diagrams
to work with the shorter set or {\it select} the single diagram to work with (\verb|SelectDiagrams()|).
The loop corrections are calculating faster when processed by single diagrams.

Then {\it matrix element} is retrieved from diagrams set (\verb|RetrieveME()|).

Before any operation with loop matrix element one should declare the {\it $N-$dimensional space} (\verb|SetNDimensionSpace()|).
Some procedures involve $N-$dimensional mode automatically for loop objects,
but most functions (arranging, simplification etc.) don't know the nature of expression they work with.
Therefore the dimensional mode should be forced.

The diagrams set and any symbolic expression may be saved and restored at next session
to refrain from job repetition (\verb|Save()|, \verb|Load()|).

\item Amplitude calculation

The amplitude evaluation (\verb|CalcAmplitude()|) is the faster way for multi-diagram process analysis.

All the particles are considered polarized.
The spinor objects are projected to abstract basis spinor.
The basis vectors are generated in numerical code to meet the non-zero denominators condition.
See sec. \ref{sec_AMPL} for method details.

\item Matrix element squaring (coupling to other)

The squaring procedure (\verb|SquareME()|) is controlled by plenty of options,
intended mostly for the performance tuning and debugging.

It basically includes {\it reduction of gamma-matrix sequences}
coupled to kinematically dependent vectors.
It reduces the number of matrices in every product to minimum.
The item-by-item {\it squaring} is followed.
For $loop \times born^{*}$ couplings the virtual integrals are involved.
See sec. \ref{sec_SQR} for details.

Amplitude calculation is fast, but its result may be more complicated than squaring expressions.
Amplitude depends on particle momenta, polarization vectors and additional basis vectors.
For unpolarized process the averaging cycle is generated in Fortran code,
and complex-numbers calculation should be performed.
The squaring result for unpolarized process is the polynomial of momenta couplings only.
Hence there is no unique answer what result is simpler for 10-15 diagrams reaction.
One should definitely use amplitude method for more than 10-15 diagrams squaring.

\item Loop diagrams analysis

The {\it tensor virtual integrals} are reduced to scalar ones (\verb|Evaluate()|, sec. \ref{sec_VI}).
The scalar coefficients of tensor integral decompositions
are also reduced to scalar integrals in the most cases (for $1-4$ point integrals).

For {\it scalar loop integrals} the tabulated values are used.
There is no reason to tabulate
%non-divergent
integrals with complicated structure.
Hence scalar integrals table contain the $A_{0}$ and $B_{0}$ integrals with different mass configuration.
It also contains a useful $D_{0}$ chiral decomposition.
Other integrals should to be resolved using \verb|LoopTools|-like \cite{LoopTools,FF} numerical programs.

The {\it renormalization} procedure is under construction now. %(sec. \ref{sec_CT}).
The counter-terms (CT) part of Lagrangian leads to CT diagrams generating.
In the nearest future the abstract renormalization constants ($\delta_m$, $\delta_f$ etc.)
will be involved and tabulated for the minimal on-shell scheme.
The automatic derivation of constants is supposed to be implemented further.
Please refer ALHEP website for implementation progress.

\item Simplification

The {\it kinematic simplification} procedure is available (\verb|KinSimplify()|, sec. \ref{sec_UTILS}).
It reduces expression using all the possible kinematic relations between momenta and invariants.
The minimization of +/* operations in huge expressions can also be performed (\verb|Minimize()|).

\item Fortran procedure creation for numerical analysis

{\it F90 or F77-syntaxes} for generated procedures are used.
Generated code can be linked to \verb|PYTHIA|, \verb|LoopTools| and any Monte-Carlo generator for numerical analysis.

%Some Fortran compilers works extremely slow with long procedures.
%Therefore Fortran functions are automatically splitted after every 30 (free parameter) lines for faster compilation.

\end{enumerate}

\section{ALHEP script language \label{sec_LANG}}

%The further features discussion is followed with adjacent script commands.
%Hence we start from the script language overview.

The script syntax is similar to C/Java languages.
Command line breaks at ";" symbol only.
Comments are marked as {\tt //} or {\tt /*...*/}.
Operands may be variables or function calls.
If no function with some title is defined, the operand is considered as variable.
The notation is case-sensitive.

The script language have no user-defined functions, classes or loop operators.
It seems to be useless in current version.

All ALHEP features are implemented as build-in functions.
The execution starts from the first line
and finishes at the end of file (or at {\tt Halt()} command).

Variable types are casted and checked automatically (in run-time mode).
No manual type specifying or casting are available.
List of ALHEP script internal types:
\vspace{-7pt}
\begin{itemize}
\item Abstract Symbolic Expression ({\it expr}).
\par Result of function operations.
Can be stored to file(\verb|Save()|) and loaded back(\verb|Load()|).
Basic operations: +,-,*. The division is implemented using \verb|Frac(a,b)| function.
Are not supposed to be inputed manually, although a few commands for manual input exist.
\item Integer Value ({\it int}).
\par Any number parameter will be casted to integer value.
Basic operations: +,-,*,$\mid$. The division is performed using \verb|Frac(a,b)| function.
No fractional values are currently available.
\item String ({\it str}).
\par String parameters are started and closed by double quotes ("string variable").
Used to specify symbol notation (e.g. momenta titles), file names, etc.
Basic operations: +(concatenate strings).
\item Set of Feynman Diagrams ({\it diagrams}).
\par Result of diagrams composing function.
%Used for outlining of matrix elements and in some operations on the diagrams list.
%Used for matrix elements outlining and in some operations on the diagrams list
Basic operations: \verb|Save()|, \verb|Load()| and \verb|SelectDiagrams()|.
Diagrams can also be converted into TeX graphics.
\item Matrix Element ({\it me}).
\par Contains symbolic expression for matrix element and information on it's use:
list of virtual momenta (integration list) etc.
\end{itemize}

%About error declarations.

The total up-to-date functions list can be found in \verb|ALHEPCommands.txt| file at project website.

\section{Initialization section \label{sec_INIT}}
\subsection{Particles \label{sec_PARTICLES}}

Particles are determined by integer number, called {\bf particle kind} (PK).
The following integer constants are defined:
\verb|ELECTRON|, \verb|MUON|, \verb|TAULEPTON|, \verb|PHOTON|, \verb|ZBOZON|, \verb|WBOZON|, 
\verb|QUARK_D|, \verb|QUARK_U|, \verb|QUARK_S|, \verb|QUARK_C|, \verb|QUARK_B|, \verb|QUARK_T|, 
\verb|NEUTRINO_ELECTRON|, \verb|NEUTRINO_MUON|, \verb|NEUTRINO_TAU|.
The ghost and scalar particles are not supposed to be external and their codes are unavailable.
Antiparticles have negative PK that is obtained by "\verb|-PK|" operation.

If kinematics is declared the particles can be secelted using the {\bf particle ID} (PID) number.
Initial particles have negeative ID (-1, -2) and final are pointed by positive numbers (1, 2, 3...).

\subsection{Kinematic selection \label{sec_KIN}}
Before any computations may be performed one needs to declare the kinematic conditions.
They are: number of initial and final state particles,
PK codes, symbols for momentum and polarization vector for every particle.
%Although the kinematics environment may further be changed, it will lead to unproper work with previously created objects.

\vspace{-3pt}
\xboxt{
\tt SetKinematics((int)N\_Initials, (int)N\_Finals, \\
\tt \qquad [(int)PK\_I, (str)momentum\_I, (str)polarization\_I, ...]);
}
\vspace{-3pt}

Here \verb|N_Initials| and \verb|N_Finals| -- numbers of initial and final particles in kinematics.
The next parameters are particle kind, momentum and polarization symbols, repeated for every particle.
For example, the $e^-(k_1,e_1) e^+(k_2,e_2) \to \mu^- (p_1,e_3) \mu^+ (p_2,e_4)$ process
should be declared as follows:

\begin{verbatim}
SetKinematics(2, 2,
              ELECTRON,"k\_1","e\_1",
              -ELECTRON,"k\_2","e\_2",
              MUON, "p\_1", "e\_3",
              -MUON, "p\_2", "e\_4");
\end{verbatim}

\subsection{Particle masses \label{sec_MASSES}}
All the particles are considered massive by default.
Particle can be declared massless using the {\tt SetMassCalcOrder} function.

\xboxt{ SetMassCalcOrder((int)PK, (int)order); }
\begin{description}
\item[\tt PK:] particle kind,
\item[\tt order:] maximum order of particle mass to be kept in calculations.
Zero value means massless calculations for specified particle.
Negative value declares the mass-exact operations.
\end{description}

The mass symbols are generated automatically and look like $m_e$, $m_W$ etc.
Hence, all the electrons in process will have the same mass symbol.
To declare unique mass symbols for specific particles the \verb|SetMassSym()| function is used.
The different masses for unique particles are often required to
involve the Breit-Wigner distribution for particles masses.

\xboxt{ SetMassSym((int)PID, (str or expr)mass); }
\begin{description}
\item[\tt PID:] particle ID in kinematics ($<0$ for initial and $>0$ for final particles),
\item[\tt mass:] new mass symbol, like \verb|"m\_X"| for $m_X$.
\end{description}

\subsection{Polarization data \label{sec_POL}}
All the particles are considered as polarized initially.
The default polarization vector symbols are set together with kinematic data (in \verb|SetKinematics()|).

To declare particle unpolarized the \verb|SetPolarized()| function is used:

\xboxt{ SetPolarized((int)PID, (int)polarized);}
\begin{description}
\item[\tt PID:] particle ID in kinematics,
\item[\tt polarized:] 1 (polarized) or 0 (unpolarized).
\end{description}

One can declare the specific polarization state (helicity value) for particles.

%The following photon ($k$,$e$) polarization definitions are available.
The photon ($k$,$e$) helicities are involved in terms of two outer physical momenta:
\begin{eqnarray} \label{e_mu}
	e^{\pm}_{\mu} = \left( (p_{+} \cdot k) {p_{-}}_{\mu} - (p_{-} \cdot k) {p_{+}}_{\mu}
	\pm i \epsilon_{\mu \alpha \beta \nu} p^{\alpha}_{+} p^{\beta}_{-} k^{\nu} \right) \slash { N_{k,p_{\pm}} }.
\end{eqnarray}
%\begin{eqnarray} \label{e_hat}
%	\hat{e}^{\pm} = -\left(\hat{k}\hat{p}_{-}\hat{p}_{+}\gamma_{\pm}
%	-\hat{p}_{-}\hat{p}_{+}\hat{k}\gamma_{\mp} \mp (p_{+}p_{-}) \hat{k} \gamma{5} \right) \slash { N_{k,p_{\pm}} }
%\end{eqnarray}

\xboxt{ SetPhotonHelicity((int)PID, (int)h, (str)"pP", (str)"pM"]); }
\begin{description}
\item[\tt PID:] photon particle ID in kinematics,
\item[\tt h:] helicity value: $\pm 1$ or $0$. Zero value clears the helicity information
\item[\tt pP, pM:] $p_{\pm}$ base vectors in (\ref{e_mu}) formula.
The $\gamma$-coupled forms of (\ref{e_mu}) (precise or chiral \cite{HBerends}) are used automatically if available.
\end{description}

One may select the transverse unpolarized photon density matrix (instead of usual $-(1/2) g_{\mu \nu}$):
\begin{eqnarray} \label{eq_hpDMp}
	\sum e_{\mu} e^{*}_{\nu} \to -g_{\mu \nu}+\frac{P_{\mu} k_{\nu}}{(P \cdot k)}+\frac{k_{\mu} P_{\nu}}{(P \cdot k)}-\frac{k_{\mu} k_{\nu} k^2}{ {(P \cdot k)}^2 }.
\end{eqnarray}

\xboxt{ SetPhotonDMBase((int)PID, (str)"P"); }
\begin{description}
\item[\tt PID:] photon particle ID in kinematics,
\item[\tt "P":] Basis vector $P$ in (\ref{eq_hpDMp}).
\end{description}

The fermion ($k$,$e$) helicity is declared using:

\xboxt{
	 SetFermionHelicity((int)PID, (int)h);
\\ SetFermionHelicity((int)PID, (srt)"h");
}
\begin{description}
\item[\tt PID:] fermion particle ID in kinematics,
\item[\tt h:] helicity value: $\pm 1$ or $0$. Zero value clears the helicity information.
Density matrix is usual: $\sum u \bar{u} \to \hat{p} \gamma_{\pm}$ (massless fermion).
\item[\tt "h":] symbol for scalar parameter in the following density matrix (massless fermion):
$\sum u \bar{u} \to (1 \slash 2) \hat{p}(1 + h \gamma_5)$.
\end{description}

Notes:
\begin{itemize}
\item \verb|SetPolarized(PID, 0)| call will also clear helicity data.
\item \verb|SetXXXHelicity(PID, 1, ...)| also sets particle as polarized (previous \verb|SetPolarized(PID,0)| call is canceled).
\item \verb|SetXXXHelicity(PID, 0, ...)| clears helicity data but does not set particle unpolarized.
\end{itemize}

\subsection{Physics selection \label{sec_PHYS}}

One can either declare the full Standard Model (in unitary or Feynman gauge) or select the parts of SM Lagrangian to be used.
The QED physics is used by default. The Feynman rules corresponds to \cite{SMDenner} paper.
%If no \verb|SetDiagramPhysicsc()| called the QED physics is supposed.
%(I.e. when script processing starts the pure QED physics is set).

\xboxt{ SetDiagramPhysics( (int)physID ); }
\begin{description}
\item[\tt physID:] physics descriptor, to be constructed from the following flags:% (using "$\mid$" operation):
\item \verb|PHYS_QED|: Pure QED interactions,
\item \verb|PHYS_Z| and \verb|PHYS_W|: Z- and W-boson vertices,
\item \verb|PHYS_MU|, \verb|PHYS_TAU|: Muons and tau leptons (and corresponding neutrinos),
%\item \verb|PHYS_TAU|: Tau lepton (and tau neutrino) physics,
\item \verb|PHYS_SCALARS|: Scalar particles including Higgs bosons,
\item \verb|PHYS_GHOSTS|: Faddeev-Popov ghosts,
\item \verb|PHYS_GAUGE_UNITARY|: Use unitary gauge, %(affects vector boson propagators and scalar particles)
\item \verb|PHYS_QUARKS|, \verb|PHYS_QUARKS_2GEN| and \verb|PHYS_QUARKS_3GEN|: $\{d,u\}$, $\{d,u,s,c\}$ and $\{d,u,s,c,b,t\}$ sets of quarks,
\item \verb|PHYS_RARE_VERICES|: vertices with 3 or more SCALAR/GHOST tales,
\item \verb|PHYS_CT|: Renormalization counter-terms (implementation in progress),
\item \verb|PHYS_SM|: full SM physics in unitary gauge (all the flags above except for \verb|PHYS_CT|),
\item \verb|PHYS_SM_Q1GEN|: the SM physics in unitary gauge with only first generation of quarks ($d,u$),
\item \verb|PHYS_ELW|: SM in Feynman gauge with no rare 3-scalar vertice,
\item \verb|PHYS_EONLY|: no muons, tau-leptons and adjacent neutrinos,
\item \verb|PHYS_NOQUARKS|: no quarks,
\item \verb|PHYS_4BOSONS_ANOMALOUS|: anomalous quartic gauge boson interactions (see \cite{AQGC}), affects on AAWW and AZWW vertices.
\end{description}

The less items are selected in Lagrangian, the faster diagram generation is performed.
%Note: the counter-term part is currently implemented for QED only.
%The weak diagrams should still be renormalized manually.

\section{Bondarev functions \label{sec_FBONDAREV}}

The Bondarev method of trace calculation is implemented according to \cite{BondarevTrace} paper.
The trace of $\gamma_\mu$-matrices product becomes much shorter in terms of $F$-functions.
The number of items for $Tr[ (1-\gamma^5) \hat{a}_1 \hat{a}_2 \cdots \hat{a}_{2n} ]$ occurs $2^{n}$.
For example, the $12$ matrices trace contains $10395$ items usually while the new method leads to $64$-items sum.

The $8$ complex functions are introduced:
\begin{eqnarray} \nonumber
	F_1 (a, b) = 2 [ (a q_{-}) (b q_{+}) - (a e_{+}) (b e_{-}) ],
	\\ \nonumber
	F_2 (a, b) = 2 [ (a e_{+}) (b q_{-}) - (a q_{-}) (b e_{+}) ],
	\\ \nonumber
	F_3 (a, b) = 2 [ (a q_{+}) (b q_{-}) - (a e_{-}) (b e_{+}) ],
	\\ \label{F_Bondarev}
	F_4 (a, b) = 2 [ (a e_{-}) (b q_{+}) - (a q_{+}) (b e_{-}) ],
	\\ \nonumber
	F_5 (a, b) = 2 [ (a q_{-}) (b q_{+}) - (a e_{-}) (b e_{+}) ],
	\\ \nonumber
	F_6 (a, b) = 2 [ (a e_{-}) (b q_{-}) - (a q_{-}) (b e_{-}) ],
	\\ \nonumber
	F_7 (a, b) = 2 [ (a q_{+}) (b q_{-}) - (a e_{+}) (b e_{-}) ],
	\\ \nonumber
	F_8 (a, b) = 2 [ (a e_{+}) (b q_{+}) - (a q_{+}) (b e_{+}) ].
\end{eqnarray}

The basis vectors $q_{\pm}$ and $e_{\pm}$ are selected as follows:
\begin{eqnarray}
 	q^{\mu}_{\pm} = {1 \over \sqrt{2} } (1,\ \pm 1 , 0, 0),
 	%\\ \nonumber
 	\qquad
 	e^{\mu}_{\pm} = {1 \over \sqrt{2} } (0,\ 0 , 1, \pm i).
\end{eqnarray}

The results for traces evaluation looks as follows:
\begin{eqnarray} \nonumber
Tr[(1 - \gamma^5) \hat{a}_1 \hat{a}_2 ] = F_1(a_1, a_2) + F_3(a_1, a_2), \quad\qquad\qquad\qquad\qquad\qquad\\ \nonumber
Tr[(1 - \gamma^5) \hat{a}_1 \hat{a}_2 \hat{a}_3 \hat{a}_4 ] = F_1(a_1, a_2) F_1(a_3, a_4) + F_2(a_1, a_2) F_4(a_3, a_4)+ \\ \nonumber
 + F_3(a_1, a_2) F_3(a_3, a_4) + F_4(a_1, a_2) F_2(a_3, a_4). \qquad\qquad\qquad
\end{eqnarray}
Please refer to \cite{BondarevTrace} paper for method details.

\xboxt{ SetParameter(PAR\_TRACES\_BONDEREV, (int)par); }
\begin{description}
\item[\tt par:] 1 or 0 -- allow or forbid Bondarev functions usage.
\end{description}

%The basis vectors $q_{\pm}$($e_{\pm}$) appear in symbolic expressions due to $F_i$-functions couplings, like
%$$F_i(a,{}^{\mu}) F_j(b,{}_{\mu}) \to 4 (a e_{\pm}) (b q_{\pm})$$.
%The ${F_i}_{\mu} {F_j}^{\mu}$ coupling takes origin from
%$Tr[..\gamma_{\mu}]Tr[..\gamma^{\mu}]$ construction, that is usually solved by Fiertz identity.
%Hence, the $q_{\pm}$($e_{\pm}$) basis vectors should not appear in result.
%
%The ALHEP internal titles for $q_{\pm}$ and $e_{\pm}$ vectors are: \verb|BqP|,\verb|BqM|,\verb|BeP| and \verb|BeM|.
%The different symbols may be set using:
%\xboxt{ SetBondarevBaseSyms( (sym or str)qP, qM, eP, eM); }
%\begin{description}
%\item[\tt qP,qM,eP,eM:] Symbols for $q_{\pm}$ and $e_{\pm}$ vectors.
%\end{description}

The numerical code for $F$-functions (\ref{F_Bondarev}) is contained in \verb|alhep_lib.F| library file.
The code is available for scalar couplings only: $F_{\mu \nu} p^{\mu} q^{\nu}$.
If vector Bondarev functions remain in result, the Fortran-generation procedure fails.
One should repeat the whole calculation without Bondarev functions in that case.

\section{Diagrams generation \label{sec_DIAGRAMS}}

The diagrams are generated after the kinematics and physics model are declared.
The Wick-theorem-based method is implemented.

The only distinct from the usual Feynman diagrams is the following:
the vertex rules have additional $1/2$ factors for every identical lines pair.
It makes two effects:
\begin{itemize}
\item[-] The crossing diagrams are usually involved if they have different topology from original.
I.e. if two external photon lines starts from single vertex, they are not crossed.
Nevertheless {\it all} the similar external lines have crossings in ALHEP.
\item[-] Some ALHEP diagrams have $2^n$ factors due to identical internal lines.
For example, the $W^{+}W^{-}\to \{\gamma\gamma WW vertex\} \to \gamma\gamma \to \{\gamma\gamma WW vertex\} \to W^{+}W^{-}$ diagram will have additional factor $4$.
The result remains correct due to $1/2$ factors at every $\gamma\gamma W^{+}W^{-}$ vertex.
\end{itemize}
%The user should not worry about this feature.

The diagrams are generated without any crossings.
The crossed graphs are added automatically during the squaring or amplitude calculation procedures.

%The matrix element for a selected process in a specific order of pertrubation theory
%is obtained from the interaction hamiltinian by virtue of Wick theorem.

% NOTE: decription below if OLD!!!
%ALHEP program generates Feynman diagrams in the several steps.
%\begin{itemize}
%\item Compose the operator sequence for pertrubation series element:
%$$\frac{(-i)^n}{n!} T \left[ \hat{H_1} \hat{H_2} \ldots \hat{H_n} \right] \to \frac{(-i)^n}{n!} T \left[ \hat{A} \hat{B} \ldots \hat{Z} \right]$$
%The hamiltonian expression is predefined for a specific physics model (see sec. \ref{sec_PHYS}).
%\item Expand chronological operator using the Wick Theorem
%$$T \left[ \hat{A} \hat{B} ... \right] \to N\left[ \hat{A} \hat{B} \ldots \right] + N\left[ {A^{(1)}} {B^{(1)}} \ldots \right] + \ldots$$
%The index here is the coupling number.
%\item Factorize the equivalent expressions ($1 \slash n!$-factor reduce).
%\item Couple to initial/final state vectors.
%\item Arrange common expression sign (the most of diagrams should have sign "+1").
%\item Restore diagrams topology.
%\item If physics model requires counterterms (CT) consideration,
%the CT-diagrams are generated by calling the same procedure for $(n-2)$ order,
%and using extended (with CT-additions) hamiltonian.
%The additional diagrams are added to the whole diagrams set.
%\end{itemize}

\xboxt{ diagrams ComposeDiagrams((int)n); }
\begin{description}
\item[\tt n:] order of diagrams to be created, $M_{X \to Y} \sim e^n$. %\mathcal{H}
\end{description}
Uses current physics and kinematics information. Returns generated diagrams set.

One may select specified diagrams into another diagrams set:

\xboxt{ diagrams SelectDiagrams( (diagrams)d, (int)i0 [, i1, i2...]); }
\begin{description}
\item[\tt d:] initial diagrams set. Remains unaffected during the procedure.
\item[\tt i0..iN:] numbers of diagrams to be selected. First diagram  is "0".
\end{description}
The new diagrams set is returned.

To retrieve matrix element from the diagram set the \verb|RetrieveME()| is used.
\xboxt{ me RetrieveME( (diagrams)d ); }
\begin{description}
\item[\tt d:] diagrams list.
\end{description}

\section{Helicity amplitudes \label{sec_AMPL}}
The amplitudes are calculated according to \cite{BondarevAmpl} paper.
%The main feature of this method is absence of phase factor between diagrams.
Every spinor in matrix element is projected to common abstract spinor:
\begin{eqnarray} \nonumber
\bar{u}_p = \frac{\bar{u}_Q u_p \bar{u}_p}{\bar{u}_Q u_p} = e^{i C} \frac{\bar{u}_Q P_p}{(Tr[P_p P_Q])^{1/2}},
\quad
u_p = \frac{u_p \bar{u}_p u_Q }{\bar{u}_p u_Q} = e^{i C} \frac{P_p u_Q}{(Tr[P_Q P_p])^{1/2}}.
\end{eqnarray}
The $e^{i C}$ factor is equal for all diagrams and may be neglected.

The projection operator $P_Q = u_p \bar{u}_p$ is choosen as follows:
\begin{itemize}
	\item $P_Q = \hat{Q} (1+\gamma_5)/2$ -- for massive external fermions,
	\item $P_Q = \hat{Q} (1+\hat{E_Q})/2$ -- if one of fermions is massless.
\end{itemize}

The value of vector $Q$ is selected arbitrary in Fortran numerical procedure.
The additional basis vector $E_Q$ (if exists) is selected meet the polarization requirements ($(E_Q.Q)=0$, $(E_Q.E_Q)=-1$).
The fractions like $1/Tr[P_Q P_p]$ may turn $1/0$ at some $Q$ and $E_Q$ values.
The denominators check procedures are generated in Fortran code
(the \verb|CODE_CHECK_DENOMINATORS| key should be used in \verb|CreateFortranProc()| call).
If the $\mid Tr[P_Q P_p] \mid > \delta$ check is failed, the another $Q$ and $E_Q$ values are generated.

\xboxt{ expr CalcAmplitude( (me)ME ); }
\begin{description}
\item[\tt ME:] matrix element retrieved from the {\it whole} diagrams set.
\end{description}

The result expression is a function of all the particle helicity vectors.
The averaging over polarization vectors is performed numerically.
The numerical averaging procedure is automatically generated in Fortran output if unpolarizaed particles are declared.

The \verb|CODE_IS_AMPLITUDE| key in \verb|CreateFortranProc()| procedure
declares expression as amplitude and leads to proper numerical code ($Ampl \times Amlp^{*}$).

\section{Matrix Element squaring \label{sec_SQR}}

%One can specify the {\it way of photon polarization implication}.
%Photon helicity vectors can be inserted in vector $\varepsilon_{\mu}$ or gamma-coupled form $\hat{\varepsilon}$,
%the hat-form can be both chiral and precise one,
%the vector form can be involved either before or after the squaring procedure.
%The calculation using the chiral $\hat{\varepsilon}$ form is usually fastest,
%but the precise form $\varepsilon_{\mu}$ is required for some processes.

The squaring procedure have the following steps:
\begin{itemize}
	\item matrix elements simplification to minimize the $\gamma$-matrices number,
	\item denominators caching to make procedure faster,
	\item item-by-item squaring (coupling to other conjugated),
	\item saving memory mechanism to avoid huge sums arranging (\verb|SQR_SAVE_MEMORY| option),
	\item virtual integrals reconstruction.
\end{itemize}

%	The integration over virtual particles phase space is than involved.
%	Here all the loop integrals appear

\xboxt{ expr SquareME((me)ME1, [(me)ME2,] [(int)flags ]); }
\begin{description}
	\item[\tt ME1:] matrix element \verb|#1|,
	\item[\tt ME2:] matrix element \verb|#2| (should be omitted for squaring),
	\item[\tt flags:] method options (defauls is "0"):
	\item \verb|SQR_CMS|: c.m.s. consideration (initial momenta are collinear).
The additional pseudo-covariant relations ($p_1.\varepsilon_2 \to 0, p_2.\varepsilon_1 \to 0$) appear that simplify work with abstract polarization vectors.
	\item \verb|SQR_NO_CROSSING_1|: do not involve crossings for ME \verb|#1|,
	\item \verb|SQR_NO_CROSSING_2|: do not involve crossings for ME \verb|#2|,
	\item \verb|SQR_MANDELSTAMS|: allow Mandelstam variable usage,
	\item \verb|SQR_PH_GAMMA_CHIRAL|: tries to involve photons helicities in short chiral $\hat{e}^{\pm}$ form (see sec. \ref{sec_POL}, \cite{HBerends}).
	\item \verb|SQR_PH_GAMMA_PRECISE|: involve precise $\hat{e}^{\pm}$ form for photon helicities.
If the polarization vectors are not coupled to $\gamma_{\mu}$, the vector form $e^{\pm}_{\mu}$ is used (see sec. \ref{sec_POL}).
%Note: if Gamma*E form for photon helicity is impossible (ME is complicated the vector representation for photon polarization vector will be used.
	\item \verb|SQR_SAVE_MEMORY|: save memory and processor time for huge matrix element squaring.
Minimizes sub-results by every $1000$ items and skips final arranging of the whole sum.
No huge sums occurs in calculation in this mode, but the result is also not minimal.
If result expression contains a sum of $10^5-10^6$ items (when expanded), the arranging time is significant and \verb|SQR_SAVE_MEMORY| flag should be involved.
If no results are calculated in reasonable time the \verb|CalcAmplitude()| (see sec. \ref{sec_AMPL}) procedure should be used.
\end{description}

If two matrix elements are given, the first will be conjugated, i.e. the result is $ME1^{*} \times ME2$.

\section{Virtual integrals operations \label{sec_VI}}

The {\it tensor virtual integrals} are reduced to scalar ones using two methods.

If tensor integral is coupled to external momentum $I_{\mu} p^{\mu}$ 
and $p$-vector can be decomposed by integral vector parameters, the {\it fast reduction} is involved.
The $D_x$ integrals for $2 \to 2$ process contain the whole basis of $4$-dimension space
and $D_x$-couplings to any external momentum can be decomposed.
It works well if all the polarization vectors are constructed in terms of external momenta.

The {\it common tensor reduction scheme} is involved elsewhere.
Tensor integrals are decomposed by the vector basis like
$$I_{\mu \nu} \left(p,q\right) = I \left(p_{\mu}p_{\nu}, q_{\nu}q_{\mu}, p_{(\mu}q_{\nu)}, g_{\mu \nu}\right).$$
The linear system for scalar coefficients is composed and solved. Implemented for $B_{j}$ and $C_{i,ij}$ integrals only.
The other scalar coefficients should be calculated numerically \cite{LoopTools}.
%The $D_{xxx}$ - integrals can be usually reduced using the first algorithm.

\xboxt{ expr Evaluate( (expr)src ); }

\verb|Evaluate|: Reduction of tensor virtual integrals to scalar ones.
The scalar coefficients in tensor VI decomposition are also evaluated (not for all integrals).
%\begin{itemize} \item[\it src:]  source expression (unaffected) \end{itemize}

\xboxt{ expr ConvertInvariantVI( (expr)src ); }

\verb|ConvertInvariantVI|: Vector parameters are substituted by scalars. I.e.:
$$C_0(k_1,k_2,m_1,m_2,m_3) \to C_0(k_1^2,(k_1-k_2)^2,k_2^2,m_1,m_2,m_3).$$
Note: $A$- and $B$- integrals are converted automatically during arrangement.
%\begin{itemize} \item[\it src:]  source expression (unaffected) \end{itemize}

\xboxt{ expr CalcScalarVI( (expr)src ); }

\verb|CalcScalarVI|: Substitutes known scalar integrals with its values.
The most of UV-divergent integrals ($A_0$,$B_0$) are substituted.
The chiral decomposition for $D_0$-integral is also applied.
The complicated integrals should be calculated numerically \cite{LoopTools}.
%\begin{itemize} \item[\it src:]  source expression (unaffected) \end{itemize}

The source expressions are unaffected in all the functions above.

\section{Regularization \label{sec_REGUL}}

The dimensional regularization scheme is implemented.
One may change the space-time dimension before every operation:

\xboxt{ SetNDimensionSpace( (int)val ); }
\begin{description} \item[\tt val:] 1 ($n$-dimensions) or 0 ($4$-dimensions space). \end{description}

\xboxt{ expr SingularArrange( (expr)src ); }

\verb|SingularArrange|: turns expression to 4-dimensional form.
Calculates $(n-4)^{i}$ factor in every item and drops out all the neglecting contributions.
%\begin{itemize} \item[\it src:]  source expression (unaffected) \end{itemize}

\xboxt{ SetDRMassPK( (int)PK ) }

\verb|SetDRMassPK|: set the particle to be used as DR mass regulator.

\begin{description}
	\item[\tt PK:] particle kind (see sec. \ref{sec_PARTICLES}). "0" value declares the default "$\mu$" DR mass.
\end{description}

%\xboxt{ } \begin{itemize} \item[\it :]  \end{itemize}

%\section{Renormalization \label{sec_RENORM}}
%
%{\it (In progeress of implementation for today, please refer website for updates)}
%
%The renormalization procedure is implemented via the counter-terms (CT) part of SM Lagrangian.
%The additional CT diagrams are generated and calculated
%in terms of abstract renormalization constants ($\delta_m$, $\delta_f$ etc.).
%The values of CT constants are tabulated (the minimal on-shell scheme is used)
%and may be inserted after the CT diagrams are calculated.
%
%{\it Note}: the counter-term part is currently implemented for QED only.
%The weak diagrams should still be renormalized manually.
%The future version of ALHEP program will definitely include the complete renormalization scheme.
%
%{\it Note}: the counter-term part is currently implemented for QED only.
%The weak diagrams should still be renormalized manually.
%The future version of ALHEP program will definitely include the complete renormalization scheme.
%
%\xboxt{ expr RestoreRenormConstants( (expr)src ) }
%\verb|RestoreRenormConstants|: Exchange \delta_X renormalization constants with their values in fixed renormalization scheme.

\section{ALHEP interfaces \label{sec_INTERFACES}}
\subsection{Fortran numerical code \label{sec_FORTRAN}}

The numerical analysis in particle physics is commonly performed
using the Fortran programming language.
Hence, we should provide the Fortran code
to meet the variety of existing Monte-Carlo generators.

To start a new Fortran file the \verb|NewFortranFile()| function is used:

\xboxt{ int NewFortranFile( (str)fn [, (int)type]); }
\begin{description}
	\item[\tt fn:] output file name
	\item[\tt type:] Fortran compiler conventions: \verb|CODE_F77| or \verb|CODE_F90|. The \verb|FORTRAN 77| conventions are presumed by default.
	\item Function returns the ID of created file.
\end{description}

Then we can add a function to FORTAN file:

\xboxt{ CreateFortranProc( (int)fID, (str)name, (expr)src [, (int)keys]); }
\begin{description}
	\item[\tt fID:] file ID returned by \verb|NewFortranFile()| call.
	\item[\tt name:] Fortran function name.
	\item[\tt src:] symbolic expression to be calculated.
	The source may be ${\mid M \mid}^2$, $d \sigma / d \Gamma$ (use \verb|CODE_IS_DIFF_CS| flag) or amplitude (\verb|CODE_IS_AMPLITUDE| flag).
	The result is always $d \sigma / d \Gamma$ calculation procedure.
	\item[\tt keys:] option flags for code generation (default is \verb|CODE_REAL8|):
	\item \verb|CODE_REAL8|: mean all the symbols in expression as \verb|REAL*8| values.
	\item \verb|CODE_COMPLEX16|: declare variables type as COMPLEX*16.
	\item \verb|CODE_IS_DIFF_CS|: the source expression is differential cross section.
	\item \verb|CODE_IS_AMPLITUDE| the source expression is amplitude and a squaring code should be generated: \verb|AMPL*DCONJG(AMPL)|.
	\item \verb|CODE_CHECK_DENOMINATORS| check denominators for zero. Used to re-generate free basis vectors in amplitude code (see sec. \ref{sec_AMPL}).
	\item \verb|CODE_LOOPTOOLS|: Create \verb|LoopTools| \cite{LoopTools} calls for virtual integrals.
	\item \verb|CODE_SEPARATE_VI| create unique title for every virtual integral function (according to parameter values).
	Do not use with \verb|CODE_LOOPTOOLS|.
	\item \verb|CODE_PYTHIA_VECTORS| retrieve vector values from \verb|PYTHIA| \cite{PYTHIA} user process \verb|PUP(I,J)| array.
	\item \verb|CODE_POWER_PARAMS| factorize and pre-calculate powers if possible.
	\item \verb|CODE_NO_4VEC_ENTRY| do not create a 4-vector entry for function.
	\item \verb|CODE_NO_CONSTANTS| do not use predefined physics constants. All variables becomes external parameters.
	\item \verb|CODE_NO_SPLIT| do not split functions by $100$ lines.
	\item \verb|CODE_NO_COMMON|: don't use CONNON-block to keep internal variables.
%	\item \verb|CODE_NO_COMMON_ARRAYS|: don't use arrays in internal CONNON-block values.
\end{description}

The scalar vector couplings, Bondarev functions (see sec. \ref{sec_FBONDAREV})
and $\varepsilon_{a b c d} p_1^a p_2^b p_3^c p_4^d$ objects are replaced with scalar parameters to be calculated once.
These functions are calculated using \verb|alhep_lib.F| library procedures.

Some compilers works extremely slow with long procedures.
Therefore Fortran functions are automatically splitted after every \verb|100| lines for faster compilation.

The internal functions functions have \verb|TEMPXXX()| notation.
To avoid problems with several ALHEP-genarated files linking
we should rename the \verb|TEMP|-prefix to make internal functions unique.

\xboxt{ SetParameterS(PAR\_STR\_FORTRAN\_TEMP, (str)prefix); }

\begin{tabular}{rl}
\verb|prefix|: & \verb|"TMP1"| or another unique prefix for temporary functions name.
\end{tabular}

To obtain the better performance of numerical calculations
ALHEP provides the mechanism for minimization of $+ \slash \times$ operations in the expression.
We recommend to invoke the \verb|Minimize()| function before Fortran generation (see sec. \ref{sec_UTILS}).

\subsubsection{PYTHIA interface \label{sec_PY}}

The \verb|PYTHIA| \cite{PYTHIA} interface is implemented in terms of \verb|UPINIT/UPEVNT| procedures
and $2 \to N$ phase-space generator.

The momenta of external particles are retrieved from PYTHIA user process event common block (\verb|PUP(I,J)|).
The order of particles in kinematics should meet the order of generated particles, and the \verb|CODE_PYTHIA_VECTORS| option should be used.

The template \verb|UPINIT/UPEVNT| procedures for \verb|ALHEP| $\to$ \verb|PYTHIA| junction are found at ALHEP website.
However one should modify them by adjusting the generated particles sequence,
including user cutting rules, using symmetries for calculation similar processes in single function etc.

The plane $2 \to N$ phase-space generator in \verb|alhep_lib.F| library file is written by V. Mossolov.
It is desirable to replace the plain phase-space generator by the adaptive one for multiparticle production process.

The more automation will be implemented in future. Please refer to ALHEP website for details.

\subsection{Mathematica}
The output interface to Mathematica Notebook \cite{Mathematica} file is basically used to view the expressions in the convenient form.
Implemented for all the symbolic objects in ALHEP.

\xboxt{
 SaveNB( (str)fn, (expr or me)val [, (str)comm ][, (int)flags ]);
	\\
 MarkNB( (str)fn [, (str)comm ][, (int)flags ]);
}
\begin{description}
	\item[\tt fn:] Mathematica output file name.
	\item[\tt val:] expression to be stored.
	\item[\tt comm:] comment text to appear in output file (\verb|""| if no comments are required).
	\item[\tt flags:] file open flags (default is "0"):
	\item 0: append to existing file and do not close it afterward,
	\item \verb|FILE_START|: delete previous, start new file and add Mathematica header,
	\item \verb|FILE_CLOSE|: add closing Mathematica block.
\end{description}
The \verb|MarkNB| function is used to add comments only.

The Mathematica program can open valid files only.
The valid \verb|x.nb| file should be {\it started} (\verb|FILE_START|) and {\it closed once} (\verb|FILE_CLOSE|).
It is convenient to insert \verb|MarkNB("x.nb","",FILE_START)| and \verb|MarkNB("x.nb","",FILE_CLOSE)| calls
to start and end of your script file.

No backward interface (\verb|Mathematica| $\to$ \verb|ALHEP|) is currently available.

\subsection{LaTeX \label{sec_LATEX}}

The LaTeX interface in ALHEP is only implemented for Feynman diagrams drawing.
The diagrams are illustrated in terms of AxoDraw \cite{AxoDraw} package.

\xboxt{
 DrawDiagrams( (diagrams)d, (str)fn [, (int)flags][, (int)draw]);
	\\
 MarkTeX( (str)fn, [, (str)comm][, (int)flags]);
}

\begin{description}
	\item[\tt d:] diagrams set.
	\item[\tt fn:] output LaTeX file name.
	\item[\tt flags:] file open flags (\verb|0|(default): append, \verb|FILE_START|: truncate old):
	\item[\tt comm:] comment text to appear in output file (\verb|""| if no comments are required).
	\item[\tt draw:] draw options (default is \verb|DD_SMALL|):
	\item \verb|DD_SMALL|: diagrams with small font captions.
	\item \verb|DD_LARGE|: diagrams with large font captions.
	\item \verb|DD_MOMENTA|: print particles momenta.
	\item \verb|DD_SWAP_TALES|: allow arbitrary order for final-state lines (option is included automatically for $2 \to N$ kinematics).
	\item \verb|DD_DONT_SWAP_TALES|: deny the \verb|DD_SWAP_TALES| option. The order of final lines is same to kinematics declaration.
\end{description}

%The complete implementation of TeX interface is the important task in future ALHEP development.

\subsection{ALHEP native save/load operations \label{sec_IO}}

The native ALHEP serialization format is XML-structured.
It may be edited outside the ALHEP for some debug purposes.

\xboxt{
 Save( (str)fn, (diagrams or expr)val);
	\\
	object Load( (str)fn );
}
\begin{description}
	\item[\tt fn:] output XML file name.
	\item[\tt val:] diagrams set or symbolic expression to be stored.
\end{description}

\section{Common algebra utilities \label{sec_UTILS}}

The following set of common symbolic operations is available:
\begin{itemize}
	\item \verb|Expand()|: expands all the brackets and arranges result.
	Works slowly with huge expressions.
	\item \verb|Arrange()|: arranges expression (makes alphabetic order in commutative sequences).
	The most of ALHEP functions performs arranging automatically and there is no need to call \verb|Arrange()| directly.
	\item \verb|Minimize()|: reduces the number of "sum-multiply" operations in expression.
	Should be used for simplification of big sums before numerical calculations.
	\item \verb|Factor()|: factorize expression.
	\item \verb|KinArrange()|: arranges expression using kinematic relations,
	\item \verb|KinSimplify()|: simplify expression using the kinematic relations.
	Works very slowly with large expressions. Mostly useful for $2 \to 2$ process.
\end{itemize}

\xboxt{
	expr Expand( (expr)src );
	\\
	expr Arrange( (expr)src );
	\\
	expr Minimize( (expr)src [, (int)flags]);
	\\
	expr Factor( (expr)src [, (int)flags]);
	\\
	expr KinArrange( (expr)src [, (int)flags]);
	\\
	expr KinSimplify( (expr) src [, (int)flags]);
}
\begin{description}
	\item[\tt src:] source expression (remains unaffected).	
	\item[\tt Minimize()] function flags (default is \verb|MIN_DEFAULT|):
	{
	\begin{description}
	\item \verb|MIN_DEFAULT| = \verb.MIN_FUNCTIONS|MIN_DENS|MIN_NUMERATORS.,
	\item \verb|MIN_FUNCTIONS|: factorize functions,
	\item \verb|MIN_DENS|: factorize denominators,
	\item \verb|MIN_NUMERATORS|: factorize numerators,
	\item \verb|MIN_NUMBERS|: factorize numbers,
	\item \verb|MIN_ALL_DENOMINATORS|: factorize all denominators $1/(a+b) + 1/x \to (x+a+b)/(x*(a+b))) $,
	\item \verb|MIN_ALL_SINGLE_DENS|: factorize single denominators (but not sums, products etc.),
	\item \verb|MIN_VERIFY|: verify result (self-check: expand result back and compare to source).	
	\end{description}
	}
	\item[\tt Factor()] function flags (default is \verb|0|):
	{
	\begin{description}
	\item \verb|FACT_NO_NUMBERS|: do not factorize numbers,
	\item \verb|FACT_NO_DENS|: do not factorize fraction denominators,
	\item \verb|FACT_ALL_DENS|: factorize all denominators,
	\item \verb|FACT_ALL_SINGLE_DENS|: factorize all single denominators (but not sums, products etc.),
	\item \verb|FACT_VERIFY|: verify result (self-check: expand result back and compare to source).
	\end{description}
	}
	\item[\tt KinArrange()] function flags (default is \verb|0|):
	{
	\begin{description}
	\item \verb|KA_MASS_EXACT|: do not truncate masses (neglecting SetMassCalcOrder() settings),
	\item \verb|KA_MANDELSTAMS|: involve Mandelstam variables (for $2 \to 2$ kinematics),
	\item \verb|KA_NO_EXPAND|: do not expand source.
	\end{description}
	}
	\item[\tt KinSimplify()] function flags (default is \verb|KS_FACTORIZE_DEFAULT|):
	{
	\begin{description}
	\item \verb.KS_FACTORIZE_DEFAULT.: factorize functions and denominators (the first two flags below),
	\item \verb|KS_FACTORIZE_FUNCTIONS|: factorize functions before simplification,
	\item \verb|KS_FACTORIZE_DENS|: factorize denominators (including partial factorization) before simplification,
	\item \verb|KS_FACTORIZE_ALL_DENS|: factorize all denominators,
	\item \verb|KS_MASS_EXACT|: do not truncate masses (neglecting SetMassCalcOrder() settings),
	\item \verb|KS_MANDELSTAMS|: involve Mandelstam variables (for $2 \to 2$ kinematics),
	\item \verb|KS_NO_EXPAND|: do not expand source (if no simplification are found).
	\end{description}
	}
\end{description}

\section{Debugging tools  \label{sec_DEBUG}}

ALHEP allows user to control the most of internal calculation flow.
The debug info is stored to \verb|debug.nb| file (critical messages will also appear in console).
The \verb|debug.ini| file contains numerical criteria for messages to be logged, the debug levels for different internal classes.
{\it Warning}: raising the \verb|debug.ini| values leads to sufficient performance drop and enormous \verb|debug.nb| file growth.

If one feels some problems with ALHEP usage, please contact author for assistance (attaching your script file).
Do not waste the time for manual debugging using \verb|debug.ini|.

%The debug levels can also be changed from the script line:
%\xboxt{ int ChangeDebugLevel( (str)id, (int)val); }
%\begin{itemize}
%	\item[\it id:] debug class ID. Please refer to original \verb|debug.ini| file for the whole set.
%	\item[\it val:] new level value.
%\end{itemize}

There are some specific commands to view internal data.
For example, the whole list of tensor virtual integrals reduction results is kept in internal {\it data storage}
and can be dumped to Mathematica file for viewing:
\begin{itemize}
	\item \verb|str ViewParticleData((int)PID)| returns the brief information on particle settings.
	Puts the information string to console and returns it also as a string variable.
	\begin{description}
		\item[\tt PID:] particle ID in kinematics.
	\end{description}
	\item \verb|ViewFeynmanRules((str)nb_file, (int)flags)| stores Feynman rules of current physics to Mathematica file.
	\item \verb|ViewTensorVITable((str)nb_file, (int)flags))| stores tensor integrals reduction table to Mathematica file.
	The VI reduction table is filled during the \verb|Evaluate()| operation.	
	\item \verb|ViewScalarVICache((str)nb_file, (int)flags))|  stores scalar loop integrals values cache.
	The scalar VI cache is filled during the \verb|CalcScalarVI()| invocation.
	\begin{description}
		\item[\tt nb\_file:] output Mathematica file name,
		\item[\tt flags:] access flags to Mathematica file: 0,\verb|FILE_START| and/or \verb|FILE_CLOSE|.
		\item Call without parameters turns output to \verb|debug.nb| file.
	\end{description}	
\end{itemize}

\section{System commands \label{sec_SYS}}

The two system commands are useful:
\begin{itemize}
	\item \verb|Halt()|: stop further script processing.
	May be used to test the first part of script and save (\verb|Save()|) internal result.
	It first part finished successfully, it may be commented (\verb|\*...*\|) and followed by loading procedure (\verb|Load()|).
	Then script execution is restarted.
	\item \verb|Timer()|: view time elapsed since the last Timer() call (from program start for first call).
\end{itemize}

\section{Examples \label{sec_EXAMPLES}}
\subsection{Amplitude for $q \bar{q} \to W^{+} W^{-} \gamma$ \label{sec_EX1}}

Let's consider the example from sec. \ref{sec_DEMO} in details.
We also extend it for anomalous quartic gauge boson interactions \cite{AQGC}.
And we don't use the Bondarev method for traces calculation this time.
Please refer to sec. \ref{sec_DEMO} for ALHEP installation notes.

We start \verb|test.al| script from output files creation:
\begin{verbatim}
nbfile = "uuWWA_MPXXX.nb";          // Mathenatica file name
MarkNB(nbfile, "", FILE_START);     // Create file
texfile = "res.tex";                // LaTeX file name
MarkTeX(texfile, "", FILE_START);   // Create file
\end{verbatim}

Then $2 \to 3$ process kinematics and physics are declared:
\begin{eqnarray}
	u(p_1,e_1) \; \bar{u}(p_2,e_2) \to \gamma(f_0,g_0) \; W^{+}(f_1,g_1) \; W^{-}(f_2,g_2).
\end{eqnarray}
\begin{verbatim}
SetKinematics(2, 3                 // 2->3 process
  ,QUARK_U,"p\_1","e\_1"           // u
  ,-QUARK_U,"p\_2","e\_2"          // u-bar
  ,WBOZON,"f\_1","g\_1"            // W{+}
  ,PHOTON,"f\_0","g\_0"            // photon
  ,-WBOZON, "f\_2", "g\_2" );      // W{-}
SetDiagramPhysics(PHYS_SM_Q1GEN|PHYS_4BOSONS_ANOMALOUS);
\end{verbatim}
We declare physics with $u$- and $d$-quarks only.
The amplitude will be summarized for all the possible internal quarks numerically.
It requires the simple replacing of quark mixing matrix in resulting Fortran code: $U^2_{ud} \to U^2_{ud}+U^2_{us}+U^2_{ub}$).

Next we declare the $u$- and $d$-quarks massless:
\begin{verbatim}
SetMassCalcOrder(QUARK_U, 0);      // consider massless
SetMassCalcOrder(QUARK_D, 0);      // consider massless
\end{verbatim}
%SetParameter(PAR_TRACES_BONDEREV, 1);

Set polarizations to "\verb|-+UUU|":
\begin{verbatim}
SetFermionHelicity(1, -1);         // u
SetFermionHelicity(2, 1);          // u-bar
\end{verbatim}

Create diagrams set and store it to LaTeX file:
\begin{verbatim}
diags = ComposeDiagrams(3);        //e^3 order
DrawDiagrams(diags, texfile);
\end{verbatim}

\begin{figure}[htb]
	\centering
		\includegraphics[width=\linewidth, height=6in, angle=0]{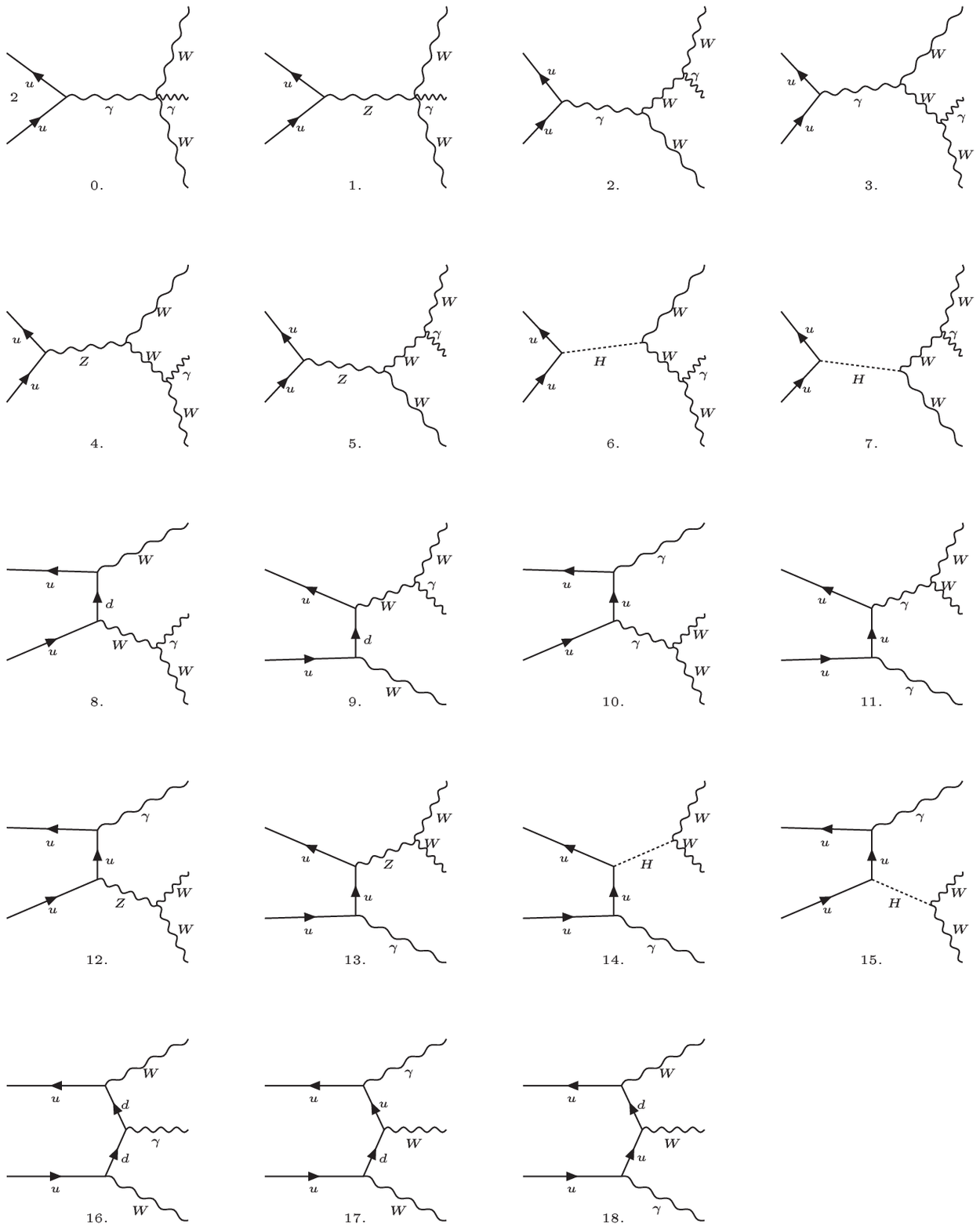}
	\caption{The diagrams generated for $u \bar{u} \to W^{-} W^{+} \gamma$ process (see {\tt res.tex} file).
	The anomalous quartic gauge boson interaction affects the first two diagrams.
	}
	\label{fig_uuWWA}
\end{figure}

Next we include the following lines:
\begin{verbatim}
Save("diags.xml",diags);           //save to XML file
//Halt();                          //stop execution
//diags = Load("diags.xml");       //load from XML file
\end{verbatim}
We can save diagrams, stop the program now and view diagrams generated.
To stop ALHEP session the \verb|Halt()| line should be uncommented.
Then we modify our script as follows:
\begin{verbatim}
/* diags = ComposeDiagrams(3);     // commented
...                                // commented
Halt(); */                         // commented
diags = Load("diags.xml");         // uncommented
\end{verbatim}
If we run the script again, it will skip the diagrams generation step and load diagrams from XML file.

Matrix element retrieval:
\begin{verbatim}
me = RetrieveME(diags);               //get matrix element
SaveNB(nbfile, me, "Matrix element"); //view
\end{verbatim}

Calculate helicity amplitude, arrange result and minimize the $+ \slash \times$ operations number:
\begin{verbatim}
ampl = CalcAmplitude(me);
SaveNB(nbfile, ampl, "Amplitude after CalcAmplitude()");
ampl = KinArrange(ampl);
SaveNB(nbfile, ampl, "Amplitude after KinArrange()");
ampl = Minimize(ampl);
SaveNB(nbfile, ampl, "Amplitude after Minimize()");
\end{verbatim}

The another breakpoint can be inserted here.
The result for amplitude is saved, the \verb|Halt| and \verb|Load| commands are commented for further use:
\begin{verbatim}
Save("ampl.xml", ampl);               // save amplitude
//MarkNB(nbfile, FILE_CLOSE); Halt(); // close NB and exit
//ampl = Load("ampl.xml");            // load amplitude
\end{verbatim}
This breakpoint allows to repeat the next Fortran creation step without recalculating of matrix element.

Let's average over final state polarizations in further numerical procedure.
Set final particles unpolarized:
\begin{verbatim}
SetPolarized(-1, 0);                  // set unpolarized
SetPolarized(-2, 0);                  // set unpolarized
SetPolarized(-3, 0);                  // set unpolarized
\end{verbatim}

The Fortran output for differential cross section:
\begin{verbatim}
SetParameterS(PAR_STR_FORTRAN_TEMP, "TMP1");
f = NewFortranFile("uuWWA.F", CODE_F77);  //f77 file
CreateFortranProc(f, "uuWWA", ampl,
    CODE_IS_AMPLITUDE|          //square amplitude
    CODE_CHECK_DENOMINATORS|    //check 1/0 limits
    CODE_COMPLEX16|             //complex values
    CODE_POWER_PARAMS|          //F(M^2) instead of F(M)
    CODE_PYTHIA_VECTORS);       //use PYTHIA PUP(I,J) vectors
\end{verbatim}
The \verb|SetParameterS| call sets the unique notation for internal variables and functions.
Please do not make it too long.
%The complex-type code is required for proper amplitude and Bondarev functions calculations.
The complex-type code is required for proper amplitude calculation.

Close Mathematica output file at the end of script:

\begin{verbatim}
MarkNB(nbfile, FILE_CLOSE);
\end{verbatim}

The execution of this script takes less than 2 minutes at 1.8GHz P4 processor.

We will not discuss the structure of generated \verb|uuWWA.F| file in details.
But some remarks should be done:
\begin{description}
	\item[Line 5:] The main function call. The following parameters are declared (order is changed here):
	All the parameters (except the \verb|kQOrig|) are of \verb|COMPLEX*16| type.
	Ones the \verb|CODE_COMPLEX16| option is set, all the real objects are treated as complex.
	\begin{description}
		\item[\tt kQOrig (INTEGER):] The ID of $u$(first)-quark in PYTHIA PUP(I,J) array. Possible values: $1$ or $2$.
		\item[\tt PAR\_a\_0, PAR\_a\_c, PAR\_a\_n, PAR\_ah\_c, PAR\_ahat\_n:]
		Anomalous quartic gauge boson interaction constants $a_0, a_c, a_n, \hat{a}_c, \hat{a}_n$ \cite{AQGC}.
		\item[\tt PAR\_CapitalLambda:]
		Scale factor $\Lambda$ for anomalous interaction \cite{AQGC}.
%		\item[\tt PAR\_Q\_d, PAR\_Q\_u:] Quark $Q_d$ and $Q_u$ charges ($2/3$ and $-1/3$).
		\item[\tt PAR\_VudP2:] Quark mixing matrix element squared ${|U_{ud}|^2}$.
		The $U^2_{ud}+U^2_{us}+U^2_{ub}$ value may be passed to summarize the whole diagrams (neglecting quarks masses).
		The numbers for mixing matrix elements may be obtained using \verb|QMIX_VAL(ID1,ID2)|, \verb|QMIX_SQR_SUM(ID)| and \verb|QMIX_PROD_SUM| functions of \verb|alhep_lib.F| library.
%		\item[\tt PAR\_Gu\_M:] $Gu_M$ constant of $Zu\bar{u}$ vertex: $i \gamma_{mu} (Gu_P \gamma_{+}+Gu_M \gamma_{-})$.
%		The numerical values are:
%		$$Gu_P = e (-2 \frac{sin_W}{3 coc_W}+\frac{1}{2 sin_W coc_W}), \quad Gu_M = -2 e \frac{sin_W}{3 cos_W}.$$
%		It useful to leave this constants as external parameters for future charge conjugation.
%		For example, the amplitude for $d\bar{d}\to W^{+}W^{-}A$ process can be obtained from $u\bar{u}\to W^{+}W^{-}A$
%		by the following exchanges:
%		$$Q_d \to -Q_u, \quad Q_u \to -Q_d, \quad Gu_P \to Gu_P' \quad Gu_M \to Gu_M' $$
%		$$\sum_x U^2_{u,x} \to \sum_x U^2_{d,x}$$
%		The minus sign before charge allows to leave the momenta order.
%		\item[\tt PAR\_m\_WP2, PAR\_m\_ZP2:] ...
	\end{description}
	\item[Line 29:] Internal COMMON-block with \verb|PAR(XX)| array.
	All the scalar couplings and other compound objects are precalculated and stored in \verb|PAR(XX)|.
	% Bondarev functions (not used here) and 
	\item[Lines 49-55:] External momenta initialization from PYTHIA \verb|PUP(I,J)| array.
	The order of external vectors is expected as follows:
		\begin{description}
		\item[\tt PUP(I,1),PUP(I,2):] initial particles.
		If \verb|kQOrig=2| the order is backward: \verb|PUP(I,2)|, \verb|PUP(I,1)|.
		\item[\tt PUP(I,3..N):] final particles in the same order as in \verb|SetKinematics()| call.
		One should modify this section (or \verb|SetKinematics()| parameters) to make the proper particles order.
		\end{description}
\begin{verbatim}
  DO 10 I=1,4
    p_1(I) = DCMPLX(PUP(I,kQ1Orig),0D0) // kQ1Orig = kQOrig
    f_0(I) = DCMPLX(PUP(I,4),0D0)
    p_2(I) = DCMPLX(PUP(I,kQ2Orig),0D0) // kQ2Orig = 3-kQOrig
    f_1(I) = DCMPLX(PUP(I,3),0D0)
    f_2(I) = DCMPLX(PUP(I,5),0D0)
  10 CONTINUE
\end{verbatim}
	\item[Lines 69-231:] Polarization averaging and basis vector generation cycle.
	For any momenta set the \verb|PAR(XXX)| array is filled. Then denominator checks and amplitude averaging are performed.
	\item[Lines 244-252, 438-444, ...] Interaction constants and particle masses definitions in sub-procedures.
	The constants can be declared as main function parameters using \verb|CODE_NO_CONSTANTS| option in \verb|CreateFortranProc| function.
\end{description}

For the complete $p \bar{p} \to W^{+} W^{-} \gamma$ analysis the following steps are required:
\begin{itemize}
\item The PYTHIA client program should be written. The template files are available at ALHEP website.
\item The another helicity configuration \verb|+-UUU| should be calculated separately.
\item The another channels $q_i \bar{q_j} \to W^+ W^- \gamma$ ($i \ne j$) should be calculated and included into generator.
\end{itemize}

\subsection{Z-boxes for $e^{-} e^{+} \to \mu^{-} \mu^{+}$ \label{sec_EX2}}

Let's calculate some box diagrams now.
Consider the following process:

\begin{eqnarray}
	e^{-}(p_1,e_1) \; e^{+}(p_2,e_2) \to \mu^{-}(f_1,g_1) \; \mu^{+}(f_2,g_2).
\end{eqnarray}

As in previous example, we start command script from files initialization:
\begin{verbatim}
nbfile = "Zbox.nb";                 // Mathenatica file name
MarkNB(nbfile, "", FILE_START);     // Create file
texfile = "res.tex";                // LaTeX file name
MarkTeX(texfile, "", FILE_START);   // Create file
\end{verbatim}

The $2 \to 2$ kinematics declaration:
\begin{verbatim}
SetKinematics(2, 2,                 // 2->2 process
  ELECTRON, "p\_1", "e\_1" ,        // e^{-}
  -ELECTRON, "p\_2", "e\_2",        // e^{+}
  MUON, "f\_1", "g\_1",             // mu^{-}
  -MUON, "f\_2", "g\_2");           // mu^{+}
\end{verbatim}

The Standard model physics in Feynman gauge (we omit quarks for faster diagrams generation):
\begin{verbatim}
SetDiagramPhysics(PHYS_ELW|PHYS_NOQUARKS);
\end{verbatim}

Set leptons massless and declare the $N$-dimensional space:
\begin{verbatim}
SetMassCalcOrder(ELECTRON, 0);      // massless electrons
SetMassCalcOrder(MUON, 0);          // massless muons
SetNDimensionSpace(1);
\end{verbatim}

Use Mandelstam variables throughout the calculation:
\begin{verbatim}
SetParameter(PAR_MANDELSTAMS,1);
\end{verbatim}

Consider unpolarized particles:
\begin{verbatim}
SetPolarized(1, 0);                 // unpolarized e^{-}
SetPolarized(2, 0);                 // unpolarized e^{+}
SetPolarized(-1, 0);                // unpolarized mu^{-}
SetPolarized(-2, 0);                // unpolarized mu^{+}
\end{verbatim}

Compose born and one-loop diagrams:
\begin{verbatim}
diags_born = ComposeDiagrams(2);    //e^2 order
DrawDiagrams(diags_born, texfile);
diags_loop = ComposeDiagrams(4);    //e^4 order
Save("diags_loop.xml",diags_loop);
//diags_loop=Load("diags_loop.xml");
DrawDiagrams(diags_loop, texfile);
\end{verbatim}
The 220 loop diagrams are created, saved to internal format and TeX file.
The \verb|ComposeDiagrams(4)| procedure takes 5-10 minutes here.
%(for \verb.PHYS_ELW|PHYS_NOQUARKS. physics)
It is convenient to comment \verb|ComposeDiagrams(4)-Save()| lines at second run and \verb|Load| loop diagrams from disk.

\begin{figure}[htb]
	\centering
		\includegraphics[width=\linewidth, height=1in, angle=0]{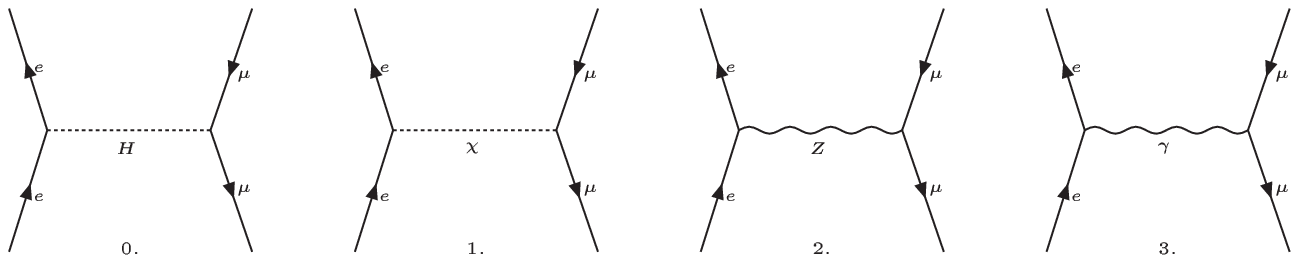}
	\caption{Born level diagrams for $e^{-}e^{+} \to \mu^{-}\mu^{+}$. }
	\label{fig_ZBoxBorn}
\end{figure}

\begin{figure}[htb]
	\centering
		\includegraphics[width=\linewidth, height=2in, angle=0]{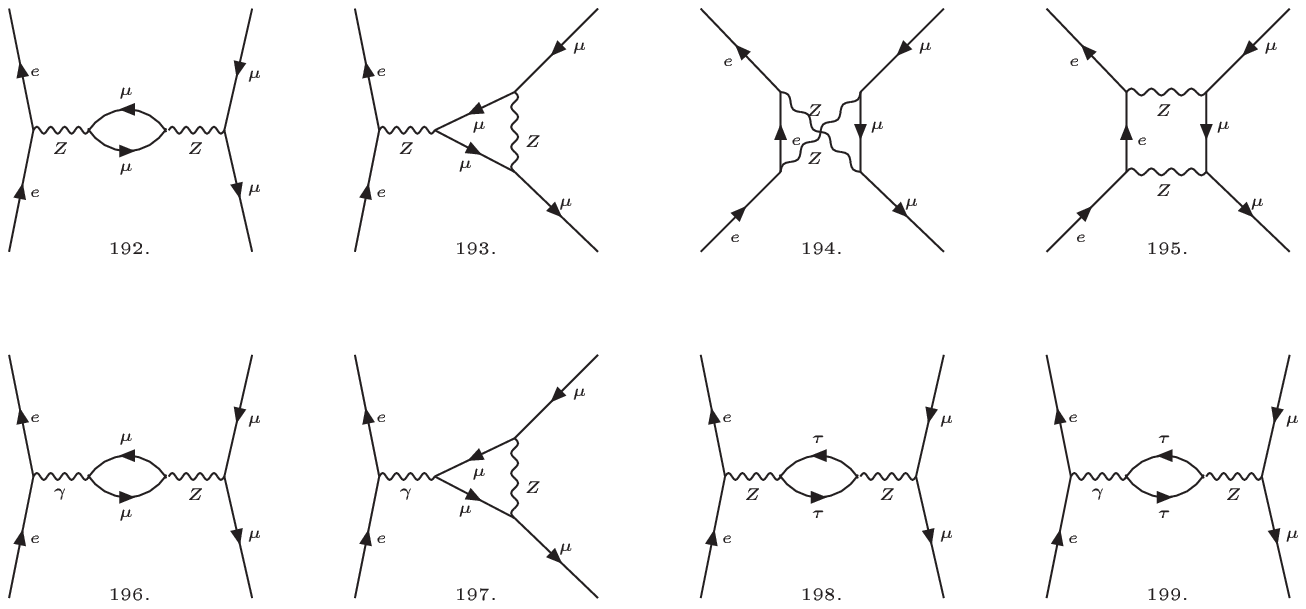}
	\caption{Part of 220 loop diagrams stored to {\tt res.tex} file }
	\label{fig_ZBox19X}
\end{figure}

Let's select the double $Z$-exchange box graphs from the whole set (194 and 195 diagrams at fig. \ref{fig_ZBox19X}):
\begin{verbatim}
diag_box = SelectDiagrams(diags_loop,194,195);
\end{verbatim}

Next we couple the loop and born matrix elements:
\begin{verbatim}
me_born = RetrieveME(diags_born);
me_box = RetrieveME(diag_box);
me_sqr = SquareME(me_box, me_born);
\end{verbatim}

The simplification procedures are not included into \verb|SquareME| implementation.
It may take much time to arrange items in huge expression.
Therefore all the simplification procedures are optional and should be called manually:
\begin{verbatim}
me_sqr = KinArrange(me_sqr);
me_sqr = KinSimplify(me_sqr);
SaveNB(nbfile, me_sqr, "squared & simplified");
\end{verbatim}

The reduction of tensor virtual integrals follows:
%The $I_{\mu} p^{\mu}$ -like objects are reduced to scalar functions:
\begin{verbatim}
me_sqr = Evaluate(me_sqr);
me_sqr = KinArrange(me_sqr);
me_sqr = KinSimplify(me_sqr);
SaveNB(nbfile, me_sqr, "VI evaluated");
\end{verbatim}

Next we convert scalar integrals to invariant-dependent form and replace with tabulated values:
\begin{verbatim}
me_sqr = ConvertInvariantVI(me_sqr);
me_sqr = CalcScalarVI(me_sqr);  // use pre-calculated values
me_sqr = KinArrange(me_sqr);
SaveNB(nbfile, me_sqr, "VI scalars ");
\end{verbatim}

Turn to $4$-dimensional space, drop out $(n-4)^i$ items and final simplification:
\begin{verbatim}
me_sqr = SingularArrange(me_sqr);
SetNDimensionSpace(0);
me_sqr = KinArrange(me_sqr);
me_sqr = KinSimplify(me_sqr);
\end{verbatim}
\begin{figure}[htb]
	\centering
	  \includegraphics[width=\linewidth, height=7in, angle=0]{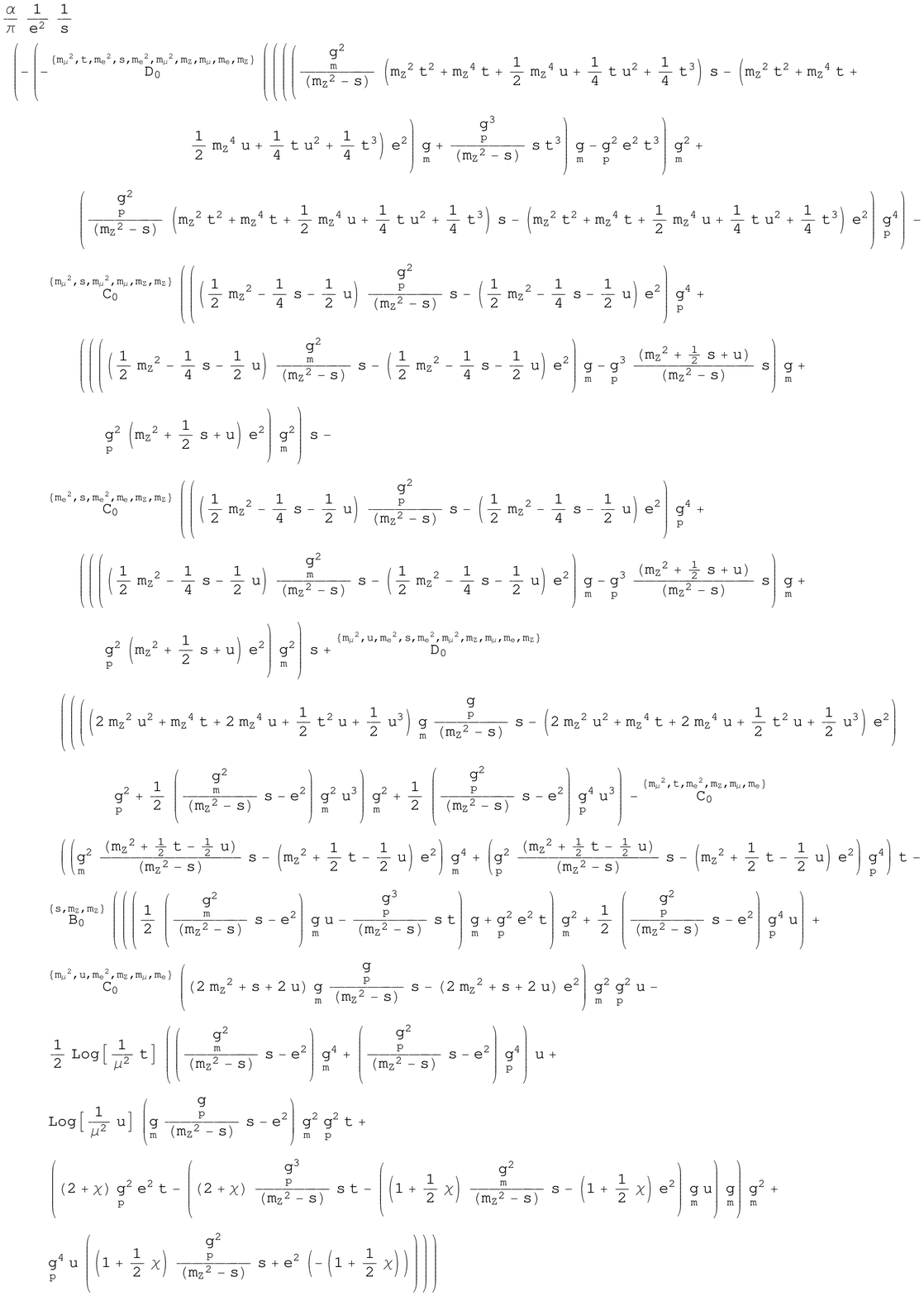}
	\caption{The result expression in {\tt Zbox.nb} file.
	This result contains UV-regulator term $\chi$, that should cancel one in $B_0(s,M_Z,M_Z)$ integral.
	It can be checked using {\tt GetUVTerm()} function. }
	\label{fig_ZBoxRes}
\end{figure}

Save result and create \verb|Fortran| code with \verb|LoopTools| \cite{LoopTools} interface:
\begin{verbatim}
Save("ZBox.xml",me_sqr);                  // save result
//me_sqr = Load("ZBox.xml");              // reload result
SaveNB(nbfile, me_sqr, "Z boxes result"); // view result
f = NewFortranFile("ZBOX.F", CODE_F77);
CreateFortranProc(f, "ZBOX", me_sqr,
                  CODE_POWER_PARAMS|CODE_LOOPTOOLS);
\end{verbatim}

View tensor integrals reduction table and close Mathematica output file:
\begin{verbatim}
ViewTensorVITable(nbfile);
MarkNB(nbfile, FILE_CLOSE);
\end{verbatim}

The script runs about 15 minutes on 1.8GHz P4 processor. The half of this time takes the \verb|ComposeDiagrams(4)| procedure.

Code remarks:
\begin{description}
	\item[Line 5:] The main function call. The 3 parameters are usual Mandelstam variables ($s, t, u$, type is complex).
%	Every part of result depends on specific pair of Mandelstam variables, but the whole expression
	The current ALHEP version does not care about interdependent parameters in Fortran output.
	And all the three Mandelstam variables may occur in parameters list.
%	and user should call \verb|ZBOX()| with 3 despite of their interdependence.
	The future versions will be saved from this trouble.
	\item[Line 14,50:] Include \verb|LoopTools| header file (\verb|"looptools.h"|). See \verb|LoopTools| manual \cite{LoopTools} for details.
	\item[Line 20,21:] Retrieve LoopTools values for UV-regulator \verb|getdelta()| and DR-mass squared \verb|getmudim()|.
\end{description}

The complete code of examples including scripts, batches and output files are available at ALHEP website.

%************************************************************************** \par

\section{Conclusions}

The new program for symbolic computations in high-energy physics is presented.
In spite of several restrictions remained in current version,
it can be useful for computation of observables in particle collision experiments.
It concerns both multiparticle production amplitudes and loop diagrams analysis.

%Examples of loop diagrams computation can be found at program website.

The nearest projects are:
\begin{itemize}
\item Bondarev functions method improvement,
\item Complete renormalization scheme for SM,
\item Complete covariant analysis of the one-loop radiative corrections
including the hard bremsstrahlung scattering contribution.
\item Arbitrary Lagrangian assignment.
\end{itemize}

%Visit the ALHEP project website for program updates.
Refer ALHEP project websites for program updates: \\
\verb|http://www.hep.by/alhep| , \\
\verb|http://cern.ch/~makarenko/alhep| .

%************************************************************************** \par

% \appendix

% \section{}
% \label{}

\end{document}